\documentstyle[12pt,epsf]{article}

\newcommand{\newc}{\newcommand}
\newc{\beq}{\begin{equation}}
\newc{\eeq}{\end{equation}}
\newc{\bea}{\begin{eqnarray}}
\newc{\eea}{\end{eqnarray}}
\newc{\kev}{\,\mbox{keV}}
\newc{\gev}{\,\mbox{GeV}}
\newc{\tev}{\,\mbox{TeV}}
\newc{\mev}{\,\mbox{MeV}}
\newc{\ev}{\,\mbox{eV}}
\newc{\gsim}{\lower.7ex\hbox{$\;\stackrel{\textstyle>}{\sim}\;$}}
\newc{\lsim}{\lower.7ex\hbox{$\;\stackrel{\textstyle<}{\sim}\;$}}
\newc{\mz}{m_Z}
\newc{\mpl}{M_{Pl}}
\newc{\obs}{{{\cal O}}}
\newc{\pb}{\,{\rm pb}}
\newc{\nn}{\nonumber}
\newc{\ie}{{\it i.e.}}
\newc{\eg}{{\it e.g.}}
\newc{\etal}{{\it et al.}}
\newc{\Z}{{\cal Z}}
\newc{\N}{{\cal N}}

\catcode`@=11
\long\def\@caption#1[#2]#3{\par\addcontentsline{\csname
  ext@#1\endcsname}{#1}{\protect\numberline{\csname
  the#1\endcsname}{\ignorespaces #2}}\begingroup
    \small
    \@parboxrestore
    \@makecaption{\csname fnum@#1\endcsname}{\ignorespaces #3}\par
  \endgroup}
\catcode`@=12

\begin{document}
\begin{titlepage}
\begin{flushright}
{\rm
IASSNS-HEP-97-136\\
UMD-PP-98-55\\
hep-ph/9712307\\
December 1997\\
}
\end{flushright}
\vskip 1cm
\begin{center}
{\Large\bf Suggested New Modes in}\\
\vskip 0.5cm
{\Large\bf  Supersymmetric Proton Decay} \\
\vskip 1cm
{\large K.S.~Babu$^{a,}$\footnote{Email: 
{\tt babu@ias.edu, $^2$pati@umdhep.umd.edu, 
$^3$wilczek@ias.edu}},
Jogesh C. Pati$^{a,b,2}$ and
Frank Wilczek$^{a,3}$\\
}
\vskip 4pt
{\large\sl $^a$School of Natural Sciences,\\ Institute for Advanced Study,\\
Princeton, NJ, USA~~08540\\}
\vskip 0.5cm
{\large \sl $^b$Department of Physics,\\ University of Maryland,\\
College Park, MD, USA 20742\\}
\end{center}
\vskip .5cm

\baselineskip = 0.22in

\begin{abstract}

We show that in supersymmetric unified theories such 
as $SO(10)$, implementation of the see--saw mechanism 
for neutrino masses introduces a new set of color triplet fields 
and thereby a new source of $d=5$ proton decay operators.  
For neutrino masses in a plausible range,
these operators are found to have the right strength to
yield observable, but not yet excluded, proton decay rates.  
The flavor structure of the new operators is distinctive.  
Proton decay modes into a charged lepton,
such as $ \ell^+ \pi^0$, $\ell^+ K^0$ and
$\ell^+ \eta$ where $\ell = e$ or $\mu$, can become prominent,
even for low or moderate values of tan$\beta \stackrel{_<}{_\sim} 10$, 
along with the $\overline{\nu} K^+$ and $\overline{\nu} \pi^+$
modes.  A distinctive feature is
the charged lepton modes involving an $e^+$ and/or
a $\mu^+$  with the ratio  
$\Gamma(\ell^+ K^0):\Gamma(\ell^+ \pi^0) \simeq
2:1$.  

\end{abstract}
\end{titlepage}
\setcounter{footnote}{0}
\setcounter{page}{1}
\setcounter{section}{0}
\setcounter{subsection}{0}
\setcounter{subsubsection}{0}


\baselineskip = 0.227in

{\bf 1.}  Proton decay, if discovered, will constitute 
impressive evidence for the placement of quarks and leptons
in common multiplets and for the unification of the
separate gauge interactions of the Standard Model \cite{pati}.  
Already, the fact that
the three gauge couplings meet at a common scale $M_X \approx
2 \times 10^{16}~GeV$ \cite{nir}, provided they are extrapolated from
their measured values in the context of supersymmetry,
supports the idea of supersymmetric unification.

Supersymmetric unified theories (GUTs), 
however, bring two new features to proton
decay: (i) First, by raising $M_X$ to a higher value as above, they
strongly suppress the gauge--boson--mediated $d=6$ proton decay
operators, for which $e^+ \pi^0$ would have been the dominant
mode.  (In the most straightforward interpretation one obtains
$\tau(p \rightarrow e^+ \pi^0)|_{d=6} \simeq 10^{36 \pm 1.5}$ yr.,
where the uncertainty reflects those from the hadronic matrix
element and from the masses of the relevant gauge bosons.)
(ii) Second, they generate $d=5$ proton decay operators \cite{sakai}
of the form $Q_i Q_j Q_k L_l/M$ in the superpotential, 
through the exchange of color triplet Higgsinos, which
are the GUT partners of the standard Higgs(ino) doublets.  
Assuming that the color triplets acquire heavy GUT--scale masses,
while the doublets remain light, these
``standard'' $d=5$ operators,
suppressed by just one power of the heavy mass and the small Yukawa
couplings, provide the dominant mechanism for proton decay in
supersymmetric GUT, with a lifetime $\tau_p \sim (10^{30}-10^{35})$ yr.
\cite{raby, murayama, babubarr, nath}.  This range is consistent with
present limits, and might be within reach of SuperKamiokande.

The flavor structure of the standard $d=5$ operators are constrained
by three factors: (a) Bose symmetry of the superfields in
$QQQL/M$, (b) color antisymmetry, and especially (c) the hierarchical
Yukawa couplings of the standard Higgs doublets.  Because of these,
it turns out that these operators lead to a strong preference for the
decay of the proton into channels involving a $\overline{\nu}$
rather than $e^+$ or (even) $\mu^+$ and those involving an $\overline{s}$
rather than a $\overline{d}$ \cite{raby}.  Thus they lead to dominant
$\overline{\nu} K^+$ and comparable $\overline{\nu} \pi^+$ modes and
in some circumstances (i.e., for large tan$\beta \stackrel{_>}{_\sim}
40$) to prominent
$\mu^+ K^0$ mode; but in all cases to highly suppressed
$e^+ \pi^0$ and $e^+ K^0$ decay modes.  For example,
in minimal $SU(5)$, with contributions only from the standard $d=5$
operators, one finds that for tan$\beta \stackrel{_<}{_\sim} 10$
(see eg., Ref. \cite{murayama, babubarr, nath}): 
\bea
\left[{\Gamma(e^+ K^0) \over\Gamma(\overline{\nu}_\mu
K^+)}\right]_{std} & \simeq & 
\left({m_u m_d \over m_c m_s {\rm sin}\theta_C}\right)^2 
{ R_{e K} \over |(1+Y_{tK})|^2} \approx 1.2 \times 10^{-7}~, \nonumber
\\
\left[{\Gamma(e^+ \pi^0) \over \Gamma(\overline{\nu}_\mu K^+)}
\right]_{std} & \simeq & \left({m_u m_d\over m_cm_s }\right)^2
{R_{e \pi} \over |(1+Y_{tK})|^2 } \approx 6 \times 10^{-8} ~,\nonumber \\
\left[{\Gamma(\mu^+ K^0) \over\Gamma(\overline{\nu}_\mu
K^+)}\right]_{std} & \simeq & 
\left({m_u \over m_c {\rm sin}^2\theta_C}\right)^2 
{ R_{\mu K} \over |(1+Y_{tK})|^2} \approx 7 \times 10^{-4} ~,\nonumber
\\
\left[{\Gamma(\mu^+ \pi^0) \over \Gamma(\overline{\nu}_\mu K^+)}
\right]_{std} & \simeq & \left({m_u \over m_c {\rm sin}\theta_C}\right)^2
{R_{\mu \pi} \over |(1+Y_{tK})|^2 } \approx 3.5 \times 10^{-4}~.
\eea
Here $R_{e\pi} \simeq R_{\mu \pi} \simeq 1.2$ and $R_{eK} \simeq
R_{\mu K} \simeq 0.12$ are the
products of the matrix element and the phase space factors for $e^+ \pi^0$
mode etc., relative to the $\overline{\nu}_\mu K^+$ mode.  The factor
$Y_{tK}$ refers to the third family contribution relative to the
second, $|Y_{tK}| \simeq |(m_t V_{td} V_{ts})/(m_cV_{cs}V_{cd})|$,
and we have neglected any flavor dependence in the squark/slepton
masses in writing Eq. (1).

The purpose of this note is to point out that there exists a
new set of color triplets and thereby plausibly a {\it new source} of $d=5$
operators, in supersymmetric unified models like $SO(10)$
\cite{so10},  which assign heavy
Majorana masses to the right--handed neutrinos to
generate light neutrino masses via the see--saw mechanism \cite{seesaw}.  
With a desirable 
pattern of the neutrino masses, in accord with the MSW solution
for the solar neutrino puzzle and $\nu_\tau$ serving as the hot 
component of dark
matter, these new $d=5$ operators are found to compete favorably with
the standard ones described above.  At the same time, the flavor
structure of the new operators, related to the neutrino masses,
appear to be rather universal, and different from the standard ones.
These new operators allow
in general, prominent or even dominant 
charged lepton decay modes of the proton--
i.e., $p \rightarrow \ell^+ \pi^0$, $\ell^+ K^0,$ and $\ell^+ \eta$,
where $\ell^+ = e^+$ or $\mu^+$, even for low values of
tan$\beta \stackrel{_<}{_\sim} 10$,  along with the 
neutrino modes $p \rightarrow \overline{\nu} K^+$
and $\overline{\nu} \pi^+$.  A distinguishing test of the new
mechanism is provided by the prominence of the charged lepton modes
involving an $e^+$ and/or a $\mu^+$, together with the prediction
$\Gamma(\ell^+ K^0): \Gamma(\ell^+ \pi^0) \simeq 2:1$.  This, as we will
discuss, can distinguish the new contributions not only from the
standard $d=5$ operators, but also from certain gauge boson mediated
effects.  

{\bf 2.}  Using standard notations for quark and lepton doublets and
also singlets, the Yukawa couplings of a
color triplet ($H_C$) and antitriplet ($H_C'$) are given 
by the superpotential 
\beq
W_{\rm Yukawa}(H_C, H_C') = F_{ij}\left[{1 \over 2} Q_i Q_j
+u_i^c \ell_j^+ \right] H_C + G_{ij}\left[Q_i L_j+u^c_i d_j^c
\right]H_C'~,
\eeq
where $i,j$ are family indices.  In the minimal $SU(5)$ model, 
$F$ and $G$ are the usual Yukawa coupling
matrices of the standard Higgs doublets with the up and the
down quarks respectively.  To allow for a different 
flavor structure in the couplings of new color triplets, we shall
keep $F$ and $G$  general.  

After integrating out $H_C$ and $H_C'$ superfields, 
the effective $\Delta B \ne 0$ superpotential is
\beq
W_{\Delta B \ne 0} = {1 \over M_C}F_{ij}G_{kl}\left[{1 \over 2}
(Q_i Q_j)(Q_k L_l) + (u_i^c \ell_j^+)(u_k^c d_l^c) \right]
\eeq
where $M_C$ is the mass of the superheavy color triplet Higgsino
($W \supset M_C H_C H_C'$).  
The $SU(3)$ and $SU(2)$ contractions in Eq. (3) are as follows:
\bea
(Q_i Q_j)(Q_k L_l) &=& \epsilon_{\alpha \beta \gamma}(u_i^\alpha
d_j^\beta-d_i^\alpha u_j^\beta)(u_k^\gamma \ell_l-d_k^\gamma
\nu_l)  \nonumber \\
(u_i^c \ell_j^+)(u_k^c d_l^c) &=& \epsilon_{\alpha \beta \gamma}
(u_i^{c \alpha}\ell_j)(u_k^{c \beta}d_l^{c \gamma})~.
\eea

In terms of component fields, Eq. (3) corresponds to a vertex
with 2 fermions and 2 scalars.  For
proton decay, the two scalars (which are heavier than the
proton) should be converted to ordinary fermions by
dressing the vertex with a wino or a gluino.  The contributions of the
gluino, which conserves flavor, turn out to be suppressed, compared to
those of the wino \cite{raby}, 
except for the case of large tan$\beta \stackrel{_>}{_\sim} 40$ (see e.g.,
Ref. \cite{babubarr}).  For the case of dominant wino contributions,
which is what we will mostly consider, only the first term
$(QQ)(QL)$ in Eq. (3) is relevant.  

{\bf 3. Neutrino masses and new dimension--5 proton decay operators:}
Now let us identify new candidates for $H_C, H_C'$ related to neutrino
masses.  Majorana masses for the right--handed neutrinos, which are 
needed to implement the see--saw mechanism, can arise in $SO(10)$ by 
utilizing the vacuum expectation value (VEV) 
of either a $\overline{\bf 126}_H$ or a $\overline{\bf 16}_H$.  In
the case of $\overline{\bf 126}_H$, one can use the renormalizable
coupling to matter multiplets (${\bf 16}_i$) of the form $f_{ij} ({\bf
16}_i {\bf 16}_j) \overline{\bf 126}_H$; while for the case of
$\overline{\bf 16}_H$, one needs to use the effective 
higher dimensional operator
$\tilde{f}_{ij}({\bf 16}_i {\bf 16}_j) (\overline{\bf 16}_H
\overline{\bf 16}_H)/M$.  Either
will generate $d=5$ proton decay operators.  We
will now discuss each, in turn.  

\underline{{\it The case of} $\overline{\bf 126}_H$}:  In this case, the
relevant standard model singlet in the $\overline{\bf 126}_H$ that
acquires a VEV has the quantum numbers of a di--neutrino ``$\nu_R
\nu_R$''.  This breaks $SO(10)$ to $SU(5)$, and as is well known, it
has the advantage that it changes $(B-L)$ by two units and thereby
automatically conserves $R$--parity \cite{bl}.  Such a symmetry 
neatly forbids potentially dangerous $d=4$ proton decay operators. 

With the $\overline{\bf 126}_H$ acquiring a VEV, 
there must exist a conjugate ${\bf 126}_H$, also acquiring VEV,
to cancel the $D$ term.  The ${\bf 126}_H$
however has no coupling to the ${\bf 16}_i$ because of
$SO(10)$.  The only relevant coupling is therefore
\beq
W_{\bf 126} ~~~=~~~ f_{ij}\left({\bf 16}_i {\bf 16}_j\right) 
\overline{\bf 126}_H~.
\eeq
Here $i,j=1,2,3$ refer to generation indices in the gauge--basis.  

The Yukawa couplings 
$f_{ij}$ may be determined (approximately) as follows.  
The light neutrino masses are given by
the see--saw formula: $m(\nu^L_i) \simeq m(\nu_i^D)^2/M_{iR}$, where
$m(\nu_i^D)$ denotes the Dirac mass of the $i$th neutrino, and
$M_{iR}$ are related to (but are not equal to) the physical Majorana masses
of $\nu_{iR}$.  $M_{iR}$ are given in terms of the matrix elements
$M_{ij} \equiv f_{ij} \left \langle \overline{\bf 126}_H \right
\rangle$ as:\footnote{This pattern emerges if one assumes a
hierarchical structure for the Dirac masses without any significant
hierarchy in the Majorana elements $M_{ij}$.} 
$M_{iR} \simeq \{M_{11}, (M_{11}M_{22}-M_{12}^2)/M_{11},
Det(M)/(M_{11}M_{22}-M_{12}^2)\}$.  The successful $SO(10)$
mass relation $m_b(M_X) = m_\tau(M_X)$ suggests that at least the
third family fermions receive their masses primarily through the
Yukawa couplings ${\bf 16}_i {\bf 16}_j {\bf 10}_H$.  This in turn
implies that $m_{\nu_\tau}^D \simeq m_t(M_X) \approx (100-120)~GeV$.  
The empirical relation $m_\mu(M_X) \approx 3m_s(M_X)$ \cite{gj}, however,
suggests that dominant contributions to the masses of the second
family comes from the Higgs component transforming as $(2,2,15)$ of
$SU(2)_L \times SU(2)_R \times SU(4)_C \equiv G_{224}$, which
contributes in the proportion $(1,1,1,-3)$ to the four colors.  Such a
Higgs component, with a VEV of the electroweak scale, may arise
effectively either from the same $\overline{\bf 126}_H$ which gives
Majorana masses to the $\nu_R^i$ (see Eq. (5)), or alternatively, and
in fact preferably, through an effective operator ${\bf 16}_i
{\bf 16}_j {\bf 10}_H \left \langle {\bf 45}_H \right \rangle/M$.  
Now ${\bf 10}_H \times {\bf 45}_H$ contains the desired submultiplet
$(2,2,15) \subset {\bf 120}$, which contributes only to the off--diagonal 
mixing (with $i,j=2,3$), as well as
a $(2,2,1)$ component.  Taking both these contributions including 
the see--saw off--diagonal mixing into account,
it can be verified that reasonable fits to the second family masses
and $V_{cb}$ can lead to $m_{\nu_\mu}^D \approx (3-12) \times
m_c(M_X) \approx (1-4)~GeV$.  Although not essential for our
arguments, guided by the masses of $u,d,$ and $e$, it seems reasonable
to take $m_{\nu_e}^D \sim (1-10)~MeV$.  

Thus, with the values of $m_{\nu_i}^D \sim \{(1-10)~MeV,~(1-4)~GeV$
and $(100-120)~GeV\}$ for $i=e,\mu,\tau$, which are motivated by the
observed pattern of masses of the quarks and the leptons, one gets,
via the see--saw formula:\footnote{These values of light neutrino masses
include a reduction of 
about 50\% owing to their running from the GUT to the electroweak scale.}  
$m(\nu^L_i) \sim \{ ({1 \over 4}-30) \times
10^{-9}, ({1 \over 4}-5) \times 10^{-3}$ and $(2-3)\}$ eV, provided
$M_{iR}$ are nearly {\it flavor universal}, within a factor of 2 to 3,
with $M_{1R} \sim M_{2R} \sim M_{3R} \sim (1-3)\times 10^{12}~GeV$.
It is interesting that this pattern of masses for the light neutrinos 
is precisely the one that goes well with the MSW solution for the 
solar neutrino puzzle, involving $(\nu_e-\nu_\mu)$ oscillations 
(which requires
$m_{\nu_\mu} \simeq (2-4)\times 10^{-3}~eV$), and with $\nu_\tau$
serving as the hot component of dark matter.

Thus we see that considerations based on quark--lepton masses as well
as neutrino masses suggest -- although they do not mandate --
a non--hierarchical pattern for
the Yukawa couplings of the $\overline{\bf 126}_H$, with a rather
universal Majorana mass $M_{iR} \sim (1-3)\times 10^{12}~GeV$.  This
contrasts with the large hierarchy exhibited in the Yukawa
couplings of the ${\bf 10}_H$ to the three families.  

In the absence of other information, it is
reasonable to take the VEVs of all relevant Higgs fields (e.g., 
$\overline{\bf 126}_H, {\bf 54}_H$ and ${\bf 45}_H$) which break $SO(10)$ to
the standard model symmetry to be nearly equal to the GUT scale,
$M_X \approx 2 \times 10^{16}~GeV$.  This also ensures 
that the simple meeting of the
gauge couplings within the MSSM framework is preserved.  With
$\left \langle \overline{\bf 126}_H \right \rangle \sim M_X$ and
$M_{iR} \approx f_{ij} \left \langle \overline{\bf 126}_H \right \rangle
\approx (1-3) \times 10^{12}~GeV$, we get $f_{ij} \approx ({1\over
2}-1)\times 10^{-4}$.  This leads
to a strength for the new $d=5$ operators (see below) of order 
$f^2/M_X \sim (10^{-8}~{\rm to}~10^{-9})/M_X$, which is
of just the right order to yield
proton decay rate in an observable range.  

\begin{figure}[t]
\centerline{
\epsfysize=1.5in
\epsfbox{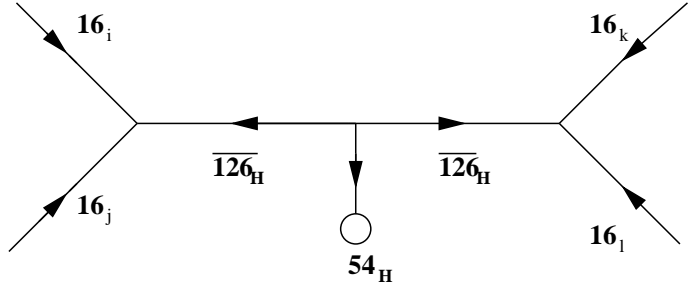}
}
\caption{Superfield diagram inducing the $QQQL/M$ operator in the
$\overline{\bf 126}_H$ option.}
\label{figone}
\end{figure}

To see the origin of the new dimension--5
proton decay operators, let us now examine the
decomposition of $\overline{\bf 126}$ under the subgroup
$G_{224} \equiv SU(2)_L \times SU(2)_R \times SU(4)_C$:
$\overline{\bf 126}= (1,3,\overline{10}) + (3,1,10) + (2,2,15)
+(1,1,6)$.  The $real$ $(1,1,6)$  component contains the
color triplet which we call $\hat{H}_C$  and an anti--triplet
($\hat{H}_C'$).  Note that Eq. (5) contains the
di--quark and lepto--quark couplings of the $\hat{H}_C$ and
$\hat{H}_C' \subset \overline{\bf 126}_H$  respectively 
(compare with Eq. (2)).  
Observe that for this case $F_{ij} = G_{ij} \equiv f_{ij} = f_{ji}$.  
Thus if a $(1,1,6).(1,1,6)$
mass term is present, there will be dimension--5
proton decay arising from the diagram shown in Fig.~\ref{figone}.  
In order to break the $SU(5)$ symmetry that is preserved
by the VEV of ${\bf 126}_H$, there must exist other Higgs
representations.  A ${\bf 45}_H + {\bf 54}_H$ is a
simple choice.  The invariant couplings
\beq
W ~~~\supset~~~\lambda \left({\bf 126}_H {\bf 126}_H {\bf 54}_H\right) + 
\overline{\lambda}\left( \overline{\bf 126}_H \overline{\bf 126}_H
{\bf 54}_H \right)
\eeq
are then allowed.  The ${\bf 54}_H$ acquires a VEV along
the $(1,1,1)$ direction under $G_{224}$.  This supplies 
the required $(1,1,6).(1,1,6)$ mass term to
induce the new effective 
dimension--5 operator of Fig.~\ref{figone}.  Since there is
also a ${\bf 126}_H \overline{\bf 126}_H$ mass term, the
two $(1,1,6)$ multiplets coming 
from ${\bf 126}_H$ and $\overline{\bf 126}_H$ will now mix with
an angle parameter $\theta$.    
This means that the
effective $M_C$ in Eq. (3) is 
$[{\rm cos}^2\theta/M_1 + {\rm sin}^2\theta/M_2]$ where 
$M_{1,2}$ are
the two mass eigenvalues of the color triplet system arising from
$\overline{\bf 126}_H$ and ${\bf 126}_H$.  

It will be seen later (Section 5) that the interaction of the ${\bf 54}_H$ in
Eq. (6) is desirable in connection with an attractive mechanism for
doublet--triplet splitting, to obtain masses for potential 
Nambu--Goldstone multiplets.
Thus there is an intricate link between the
neutrino masses, doublet--triplet splitting and the
proton decay operators in this case.

Only the $(1,1,6)$ component of the $\overline{\bf 126}_H$ 
contributes to dimension--5 proton decay operator.  
It is easy to verify that although the $(1,3,\overline{10})$ and
$(3,1,10)$--components of $\overline{\bf 126}_H$ contain
color triplets, they do not mediate proton decay.  This is a
consequence of $SO(10)$, which prevents the Yukawa couplings of the
${\bf 126}_H$ to the ${\bf 16}_i$.  

\underline{{\it The case of} $\overline{\bf 16}_H$}:  In this case, 
the relevant standard model singlet that acquires a VEV has the 
quantum numbers of
a neutrino ``$\nu_R$''.  This too breaks $SO(10)$ to $SU(5)$, but it
changes $(B-L)$ by one unit.  One can define a $Z_2$
discrete symmetry under which ${\bf 16}_i$'s are odd, but all other
multiplets (including $\overline{\bf 16}_H$ and ${\bf 10}_H$) are
even.  This symmetry can serve as $R$--parity.  

Now, with $\overline{\bf 16}_H$ acquiring a VEV, one needs a
${\bf 16}_H$, acquiring the same VEV, to cancel the $D$--term.
$SO(10)$--invariance, together with the $Z_2$ symmetry (under which 
${\bf 16}_H$ is even), now allows the superpotential terms 
\bea 
W_{\bf 16} = {1 \over M}\tilde{f}_{ij}^a  \left({\bf 16}_i 
{\bf 16}_j\right)_a 
\left(\overline{\bf 16}_H \overline{\bf 16}_H\right)_a
+ {1 \over M} \tilde{g}_{ij}\left( {\bf 16}_i 
{\bf 16}_j\right) \left( {\bf 16}_H {\bf 16}_H\right) + M_{16}
{\bf 16}_H \overline{\bf 16}_H~.
\eea
Here $a$ in the first term refers to the two possible $SO(10)$--contractions.  

While the first two non--renormalizable 
terms in Eq. (7) might be taken as quasi--fundamental, to be cut off
by gravity or string effects at short distances, it
is interesting to examine their possible origins through renormalizable 
operators.  A simple way to generate the first term in Eq. (7) that
induces the Majorana masses of the $\nu_{iR}$, is via the couplings
${\bf 16}_i {\bf 45}_j \overline{\bf 16}_H + M_{ij} {\bf 45}_i
{\bf 45}_j$.  With this coupling alone, which appears to be almost 
inevitable to induce neutrino masses, there are new contributions to
$d=5$ proton decay.  
\begin{figure}[t]
\centerline{
\epsfysize=1.5in
\epsfbox{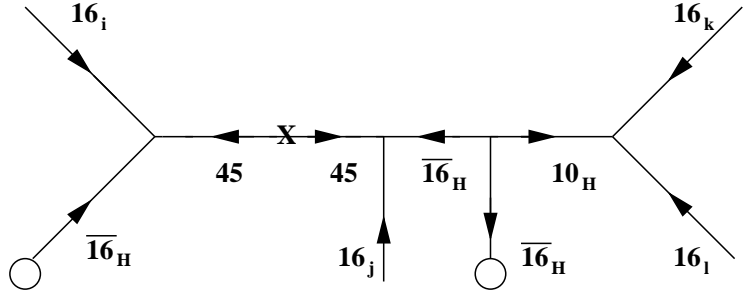}
}
\caption{Dimension 5 proton decay operator involving one
neutrino mass vertex and one standard color triplet vertex.}
\label{figtwo}
\end{figure}

\begin{figure}[t]
\centerline{
\epsfysize=1.5in
\epsfbox{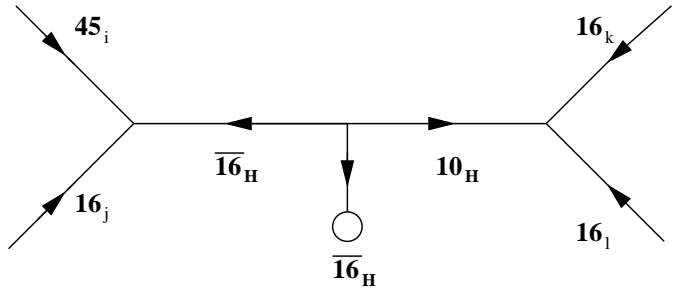}
}
\caption{Another $d=5$ proton decay operator from one
neutrino mass vertex and one standard color triplet vertex.}
\label{figthree}
\end{figure}
The relevant diagrams are shown in Figures 2 and 3.  
In Fig. 2, one of the vertices arises from the effective neutrino mass
operator, Eq. (7), while the other vertex is the 
standard operator  ${\bf 16}_i {\bf 16}_j
{\bf 10}_H$ proportional to the down quark mass matrix.  The
coupling $\overline{\bf 16}_H \overline{\bf 16}_H {\bf 10}_H$
is allowed by all the symmetries, and is also compatible with the
doublet--triplet splitting mechanism.  This term
may be desirable since it modifies the relation tan$\beta = m_t/m_b$
that often occurs in $SO(10)$ models.  (tan$\beta \sim
m_t/m_b$ is generally problematic for the standard $d=5$ proton decay.)  
This modification arises because the $Y=-1/2$ light Higgs doublet, 
will now be partly from the ${\bf 10}_H$ and partly from the ${\bf 16}_H$ (see
eg., Ref. \cite{bbfermion}).  The up--down symmetry
preserved by the usual ${\bf 16}_i {\bf 16}_j {\bf 10}_H$ Yukawa couplings 
will now be broken, resulting in tan$\beta \neq m_t/m_b$.    
Fig. 3 makes
use of the fact that the light fermions are not entirely in the ${\bf
16}_i$, but are also contained in ${\bf 45}_i$, due to the mixing from
${\bf 16}_i {\bf 45}_j \overline{\bf 16}_H$.  

Analogous to the case of $\left \langle \overline{\bf 126}_H
\right \rangle$, neutrino masses would determine $\tilde{f}_{ij}$
(approximately) by noting that $M_{iR} \approx \tilde{f}_{ij}
\left \langle \overline{\bf 16}_H \right \rangle^2/M \approx (1-3)
\times 10^{12}~GeV$.  Thus we find
$\tilde{f}_{ij} \left \langle \overline{\bf 16}_H \right \rangle/M
\sim ({1 \over 2}-1)\times 10^{-4}$, with no strong hierarchy in 
its elements.  Consequently  
the new contributions from these terms allow charged lepton decay
modes of the proton to
become prominent (see below).

The ${\bf 16}_i {\bf 16}_j{\bf 16}_H {\bf 16}_H$ term in
Eq. (7) is not directly related to neutrino masses.  However, it is
similar in form to the first term of
Eq. (7).  It
has been used in the past
to induce realistic fermion masses and mixings in 
$SO(10)$ \cite{bbfermion}.  Indeed, note that
with a single ${\bf 10}_H$ coupling to ${\bf 16}_i{\bf 16}_j$, the
up and down matrices are proportional and the CKM matrix reduces 
to the identity.
To correct these bad $SO(10)$ relations, one needs some additional
contributions to the mass matrices.  A simple solution is to induce
the effective operator ${\bf 16}_i{\bf 16}_j{\bf 16}_H {\bf 16}_H$
through the (renormalizable)
couplings ${\bf 16}_i {\bf 10}_j {\bf 16}_H + M_{ij}{\bf
10}_i {\bf 10}_j$ involving superheavy ${\bf 10}_i$.  
Then if the mixing term $\overline{\bf 16}_H 
\overline{\bf 16}_H {\bf 10}_H$ is also present, so that the $SU(2)_L$ 
doublet from the ${\bf 16}_H$ acquires a VEV, 
the proportionality relations will be corrected 
and  non--zero CKM mixings will be induced.  
From a fit to the masses and the CKM mixing angles, 
one finds that the matrix elements 
$\tilde{g}_{ij}$ are all of the same order, of the order of the
strange quark Yukawa coupling, within a factor of 10, 
$\tilde{g}_{ij}\sim (10^{-3}-10^{-4})$.  These
terms would still leave the bad relations of minimal $SU(5)$~ $m_s(M_X) =
m_\mu(M_X)$ and $m_d(M_X) = m_e(M_X)$. 
\begin{figure}[t]
\centerline{
\epsfysize=1.5in
\epsfbox{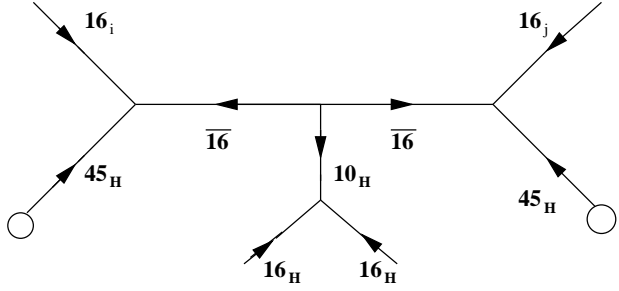}
}
\caption{Possible contribution to 
${\bf 16}_i{\bf 16}_j{\bf 16}_H{\bf 16}_H$ operator relevant to proton
decay and fermion masses.}
\label{figfour}
\end{figure}
One possible way to correct
them, while simultaneously inducing the CKM mixing angles, is to utilize
the couplings ${\bf 16}_i \overline{\bf 16}~ {\bf 45}_H + 
\overline{\bf 16}_i \overline{\bf 16}~ {\bf 10}_H+ {\bf 16}_H
{\bf 16}_H {\bf 10}_H$ involving vector--like
pairs of matter multiplets in ${\bf 16}+\overline{\bf 16}$ (see Fig. 4).  
This would induce an effective
operator $({\bf 16}_i {\bf 16}_j) ({\bf 16}_H {\bf 16}_H) ({\bf 45}_H
{\bf 45}_H)$, which can also serve the purpose of the $\tilde{g}_{ij}$
terms in Eq. (7).   In this case, again, one
sees from a fit to the quark and lepton masses and mixings, 
that the effective couplings $\tilde{g}_{ij}$ are
nearly universal and of order the strange quark Yukawa coupling.  

From all this, it seems natural to assume $\tilde{g}_{ij} \sim 
\tilde{f}_{ij}$.  The $d=5$ proton decay amplitude 
(see below) then 
turns out to have the right strength, as in the case of 
$\overline{\bf 126}_H$, to yield observable rates.

Let us focus on the proton decay operators  arising from  
utilizing all three terms of Eq. (7).  The
$SU(5) \times U(1)_X$ decomposition relevant to Eq. (7) is:   ${\bf 16} =
{\bf 1}^{-5} + \overline{{\bf 5}}^3 + {\bf 10}^{-1}$ where
the superscript indicates the $U(1)_X$ charges.  So the first term in 
Eq. (7) contains the terms
\beq
\left({\bf 1}_i^{-5} {\bf 1}_j^{-5}\right) \left({\bf 1}_H^5 {\bf
1}_H^5 \right) 
+ \left({\bf 10}_i^{-1} {\bf 10}_j^{-1}\right)
\left( {\bf 5}_H^{-3} {\bf 1}_H^5\right)
\eeq
while the second term contains
\beq 
\left(\overline{\bf 5}_i^3 \overline{\bf 5}_j^3\right) 
\left({\bf 10}_H^{-1}
{\bf 1}_H^{-5}\right) + \left(\overline{\bf 5}_i^3 {\bf 10}_j^{-1}\right)
\left(\overline{\bf 5}_H^3 {\bf 1}_H^{-5}\right)~.
\eeq
Once the ${\bf 1}_H^5$ from the $\overline{\bf 16}_H$ and
the ${\bf 1}_H^{-5}$ from the ${\bf 16}_H$ acquire (equal)
GUT--scale VEVs, the first term in Eq. (8) will induce 
superheavy Majorana masses for the $\nu_R$. The second term in
Eq. (8) along with the second term in Eq. (9) will lead to
dimension--5 proton decay.  (Recall that the ${\bf 5}_H^{-3}+
\overline{\bf 5}_H^{3}$ do not belong to the Nambu--Goldstone
supermultiplet in the $SO(10)/SU(5)$ coset space.) 
The effective operator is given by
Eq. (3), with the identification of $F$ and $G$ matrices
with the $\tilde{f}\left \langle {\bf 1}_H^5 \right \rangle/M$ and 
$\tilde{g} \left \langle {\bf 1}_H^{-5} \right \rangle/M$
matrices of Eq. (7) respectively.  (The first term
in Eq. (9) does not lead to proton decay.)  As in the case of
$\overline{\bf 126}_H$, the coupling matrices
$F$ and $G$ are now non--hierarchical. This turns out to be
very significant for proton decay.

{\bf 4. Proton decay Modes:}
We can write down the effective $\Delta B \ne 0$ four--fermion
interactions that arise after $\tilde{W}^\pm$ dressing of the operator
in Eq. (3).  We do this in the mass eigenbasis of the quarks
and the leptons.  First define the gauge interaction eigenstates
for the down quarks $d'_i = V_{ij} d_j$ and for the neutrinos
$\nu'_i = V^{\ell^*}_{ji} \nu_j$ with $V$ and $V^\ell$ being the
CKM matrix in the quark sector and the lepton sector.  Then the
effective $\Delta B \ne 0$ four--fermion interaction is
\bea
{\cal L}_{\Delta B \ne 0} &=& 
{1 \over M_C} {\alpha_2 \over 4 \pi} \hat{F}_{ij}
\hat{G}_{kl} \epsilon_{\alpha \beta \gamma} \nonumber \\
& \times & [(u_i^\alpha d_j^{\prime\beta})(d_k^{\prime\gamma} \nu_l')
\left(f(d'_i, u_j) + f(u_k, \ell_l^-) \right) \nonumber \\
& + & (d_i^{\prime \alpha} u_j^\beta)(u_k^\gamma \ell_l^-)
\left(f(u_i, d'_j)+f(d'_k, \nu_l')\right) \nonumber \\
& + & (d_i^{\prime\alpha} u_k^\beta)(d_j^{\prime\gamma} \nu'_l) \left(
f(u_i, d'_k)+f(u_j, \ell^-_l)\right) \nonumber \\
& + & (u_i^\alpha d_k^{\prime\beta})(u_j^\gamma \ell_l^-) \left(
f(d'_i, u_k) + f(d'_j,\nu'_l)\right) ]~.
\eea
Here the color indices $(\alpha, \beta, \gamma)$ and the
flavor indices $(i,j,k,l)$ are understood to be summed over,
and the fermion fields paired together in parentheses are
spin--contracted to singlets.  $f$ is a loop integral, with
magnitude of $M_{\rm SUSY}^{-1}$, defined as
\beq
f(a,b) = {M_{\tilde{W}} \over m_{\tilde{a}}^2-m_{\tilde{b}}^2 }
\left( {m_{\tilde{a}}^2 \over m_{\tilde{a}}^2 - m_{\tilde{W}}^2 }
{\rm ln} {m^2_{\tilde{a}} \over m^2_{\tilde{W}} } - [a \rightarrow b]
\right)~.
\eeq
Note that all the matter fields in Eq. (10) belong to weak isodoublets.
Traditionally it has been argued that the third term 
dominates; one of our main points here has been to emphasize that this
need not be so.  

Since Eq. (10) is in the mass eigenbasis, the Yukawa couplings
$\hat{F}_{ij}$ and $\hat{G}_{ij}$ are not the same as $F_{ij}$
and $G_{ij}$ of Eq. (2).  $\hat{F}$ remains symmetric, it is
related to $F$ by $\hat{F} = V_u^T F V_u$ while $\hat{G}
= (V_u^T G V_u)V'$, where $V_u$ is the unitary matrix
that rotates the left--handed up quarks in going to the
mass eigenbasis.  $V'$ is another unitary matrix that parameterizes
the mismatch between the up quark and the charged lepton
mass matrices \cite{jarlskog}, $V' = V_u^{\dagger} V_\ell$.  Note that
$\hat{G}$ is not symmetric.  

The proton decay rate and branching ratios can be obtained
from Eq. (10).  As we have emphasized, 
several considerations suggest the matrices
$\hat{F}$ and $\hat{G}$ may not be hierarchical,
so proton decay into alternative flavor 
modes could a priori have similar rates.
Consider first the decay into charged leptons.  The relevant
interactions are the second and the fourth terms in Eq. (10).
In the second term, we must put $j=k=1$ since the operator has
to have only $u$ quarks.  Flavor antisymmetry then requires
that $i=2$ or $3$.  Similarly in the fourth term, $i=j=1,
k=2,3$.  Noting that $d_2'\simeq V_{cd} d + V_{cs} s$
and $d_3' \simeq V_{td} d + V_{ts} s$ for proton decay, the
amplitude for charged lepton decay can be written as
\bea
A(p \rightarrow  \ell^+ )  \propto  
\{(u^\alpha d^\beta)(u^\gamma \ell^-_l) \left[V_{cd}\left(\hat{G}_{1l}
\hat{F}_{21} - \hat{G}_{2l}\hat{F}_{11}\right) + V_{td}
\left(\hat{G}_{1l}\hat{F}_{31}-\hat{G}_{3l}\hat{F}_{11} \right)
\right] \nonumber \\
 +  (u^\alpha s^\beta)(u^\gamma \ell^-_l) \left[V_{cs} \left(
\hat{G}_{1l}\hat{F}_{21}-\hat{G}_{2l}\hat{F}_{11}\right) +
V_{ts}\left(\hat{G}_{1l}\hat{F}_{31}-\hat{G}_{3l}\hat{F}_{11}
\right)\right] \}~.
\eea
Now if the matrices $\hat{F}$ and $\hat{G}$ have no strong
hierarchy in their flavor--dependence, 
then the terms proportional to $V_{td}$ and $V_{ts}$
can be ignored in Eq. (12).  The error introduced is only
of order $\lambda_C^2 \simeq 1/20$, 
$(\lambda_C \equiv {\rm sin}\theta_C \simeq 0.22)$.  
This observation leads to
predictions for the branching ratios of certain (in general) prominent
modes, which are independent of
the flavor structure in $\hat{F}, \hat{G}$:
\beq
{\Gamma(p \rightarrow \ell^+ \pi^0) \over \Gamma(p \rightarrow \ell^+ K^0) } 
\simeq {\rm sin}^2\theta_C \left(1- {m_K^2 \over m_p^2} \right)^{-2} R~.
\eeq
Here $\ell^+ = e^+$ or $\mu^+$ and $R$ is the ratio of the two relevant
hadronic matrix element--squared.
The chiral Lagrangian estimate for $R$ given in Ref. \cite{murayama} is
\beq
R = {\left|(1+D+F) \right|^2 \over 
2\left|1-{m_p \over m_B}(D-F)\right|^2} ~.
\eeq
Using $D=0.81, F=0.44$ for the chiral Lagrangian factors and with
$m_B = m_\Lambda = 1150~MeV$ (as in \cite{murayama}), 
we obtain $R \simeq 5$.  Thus 
\beq
\Gamma(p \rightarrow  \ell^+\pi^0)/\Gamma(p \rightarrow \ell^+K^0)
\simeq 0.5~.  
\eeq

Another interesting mode is $p \rightarrow \ell^+ \eta$ for which
one has 
\beq
{\Gamma(p \rightarrow  \ell^+ \eta) \over \Gamma(p \rightarrow 
\ell^+ \pi^0) } = \left(1-{m_\eta^2 \over m_p^2}\right)^2
3{\left|1-{1 \over 3}(D-3F)\right|^2 \over \left|(1+D+F)\right|^2 }
 \simeq 0.35~.
\eeq

While all charged lepton modes are expected to have
similar rates, the ratios such as
$\Gamma(p\rightarrow  e^+ \pi^0)/\Gamma(p \rightarrow  \mu^+ \pi^0)$
cannot be predicted quantitatively since they are
sensitive to the flavor structure of $\hat{F}$ and $\hat{G}$:
\beq
{\Gamma(p \rightarrow e^+ \pi^0) \over \Gamma(p \rightarrow \mu^+ \pi^0)}
\simeq \left|{(\hat{G}_{11} \hat{F}_{21} - \hat{G}_{21}\hat{F}_{11}) \over
(\hat{G}_{12}\hat{F}_{21} - \hat{G}_{22} \hat{F}_{11}) }\right|^2~.
\eeq
If one uses $\overline{\bf 16}_H$ to generate the $\nu_R$
Majorana masses, the matrices $\hat{G}$
and $\hat{F}$ are independent, so the ratio in Eq. (17) 
is in general expected to be of order
unity.  However, if a single  $\overline{\bf 126}_H$ is used for this
purpose, one
has the asymptotic relation $\hat{G} = \hat{F}V'$.  The
flavor--dependent renormalization of this relation is small.  
If in addition 
the off--diagonal entries in the Jarlskog matrix $V'$ are small, then
one has a cancellation
in the amplitude for $e^+ \pi^0$.  In this case, 
one would expect the $\mu^+ \pi^0$ mode to dominate
over the $e^+ \pi^0$ mode.
Note that in the $\overline{\bf 16}_H$ option, however, since
$\hat{G}$ and $\hat{F}$ are independent, the two modes are
expected to be comparable even if $V'$ has small off--diagonal
entries.  

In the case where $\overline{\bf 16}_H$ is used to generate the see--saw
neutrino masses, as noted earlier, new $d=5$ operators arise simply
from the first term in Eq. (7).  Then the factors $\hat{F}_{ij}$, being
related to the $\nu_{iR}$ masses, are non--hierarchical, while the
factors $\hat{G}_{ij} \simeq V_{ij}^* m_j^d$ exhibit a hierarchy.  The
charged lepton modes become prominent even in this case.  For the
amplitudes, one obtains: $A(\mu^+ K^0) \propto
(\hat{F}_{11}-{\rm sin}\theta_C \hat{F}_{21}) m_s$, while 
$A(\mu^+ \pi^0) \propto {\rm sin}\theta_C (\hat{F}_{11} - 
{\rm sin}\theta_C \hat{F}_{21})m_s$, where terms of order
$\lambda_C^2 \simeq 1/20$ have been dropped.  Again one finds that
$\Gamma(\mu^+ K^0):\Gamma(\mu^+ \pi^0) \simeq 2:1$, as in Eq. (15).  
A similar remark applies to $\Gamma(e^+ K^0): \Gamma(e^+ \pi^0)$.
Now, however, $A(e^+ K^0) \propto (\hat{F}_{21} + {\rm sin}\theta_C
\hat{F}_{11})m_d$, so that $\Gamma(\mu^+ K^0)$ is expected to be 
considerably larger than $\Gamma(e^+ K^0)$ (barring fortuitous
cancellations).  

It is worth noting that, independent of the relative importance of 
positron modes, the $\mu^+ \pi^0$ and $\mu^+ K^0$ modes arising
through the new $d=5$ operators can still compete favorably with or
supersede the $\overline{\nu}K^+$ modes, for even small or moderate
values of tan$\beta \stackrel{_<}{_\sim} 10$.  
By contrast, for the standard $d=5$ operator, 
the $\mu^+ \pi^0$ and
$\mu^+ K^0$ modes can be prominent
only for very large tan$\beta
\stackrel{_>}{_\sim} 40$ through gluino dressing.  Thus a
study of these decay modes and determination of tan$\beta$ could
distinguish between the standard and the new $d=5$ operators,
even for the case of $\overline{\bf 126}_H$.  

The decays $p \rightarrow \overline{\nu} K^+$ and $p\rightarrow
\overline{\nu} \pi^+$ from the new dimension--5 operators
have rates comparable with the charged lepton modes.  The
expectation for $\Gamma(p \rightarrow  \overline{\nu} K^+)/
\Gamma(p \rightarrow  \overline{\nu} \pi^+)$ is  similar to 
but not exactly the same as the
case of minimal supersymmetric $SU(5)$.  The flavor structure
relevant for the $\overline{\nu}$ decay is given by (in the limit of
neglecting terms of order $\stackrel{_<}{_\sim}\lambda_C^2 \simeq 1/20$):
\bea
A(p \rightarrow \overline{\nu}_\ell)  \propto  \{(u^\alpha d^\beta)(s^\gamma
\nu_\ell)[V_{ud}V_{cs}(\hat{F}_{11}\hat{G}_{2\ell}-\hat{F}_{12}\hat{G}_{1\ell})
-V_{cd}V_{cs}(\hat{F}_{22}\hat{G}_{1\ell}-\hat{F}_{12}\hat{G}_{2\ell})]
\nonumber \\
- ~(u^\alpha s^\beta)(d^\gamma \nu_\ell)
[V_{cd}V_{cs}(\hat{F}_{22}\hat{G}_{1\ell}-\hat{F}_{12}\hat{G}_{2\ell})
-V_{us}V_{cd}(\hat{F}_{11}\hat{G}_{2\ell}-\hat{F}_{12}\hat{G}_{1\ell}]
\nonumber \\
+ ~(u^\alpha d^\beta)(d^\gamma\nu_\ell)[V_{ud}V_{cd}(\hat{F}_{11}
\hat{G}_{2\ell}-
\hat{F}_{12}\hat{G}_{1\ell}) -V_{cd}^2(\hat{F}_{22}\hat{G}_{1\ell}
-\hat{F}_{12}\hat{G}_{2\ell})]\}.
\eea  
Note that in Eq. (18), the $V_{ud}V_{cs}$ term (the first term) has no
mixing angle suppression, but all the remaining terms are suppressed by
factors of at least $\lambda_C$.  It turns out that the matrix element
for the $(u^\alpha s^\beta)(d^\gamma \nu_\ell)$ is suppressed
by a factor $\simeq \lambda_C$, compared to the $(u^\alpha
d^\beta)(s^\gamma\nu_\ell)$  term, leading to 
an overall suppression factor of $\sim \lambda_C^2$ in the terms in
the second line.  Neglecting these subleading terms one finds
\beq
{ \Gamma(\overline{\nu} \pi^+) \over \Gamma(\overline{\nu} K^+) } 
\simeq {\rm sin}^2\theta_C \left(1-{m_p^2 \over m_K^2}\right)^{-2}
{\left|1+D+F\right|^2 \over \left| 1+ {m_p \over 3m_B} (D+3F)\right|^2}
\simeq {1 \over 5}~.
\eeq
This prediction  holds if the contributions
from the new $d=5$ operators dominate.   
The analogous number for the standard
$d=5$ operators in minimal $SU(5)$ is near
${1 \over 8}$, but there is considerable  uncertainty
in this case owing to
possible cancellation between the second and the third generation
contributions.  It is difficult to make quantitative estimate of
ratios such as $\Gamma(\ell^+ K^0)/\Gamma(\overline{\nu} K^+)$ etc.,
arising from the new $d=5$ operators, since there is some flavor 
dependence.  Comparing the flavor structure in Eq. (12) and
Eq. (18), one infers that the rates for both these modes are
similar, with the $\overline{\nu} K^+$ mode slightly preferred
over the $\ell^+ K^0$ mode, owing to a matrix element enhancement
and the availability of final states with all three neutrino flavors.  
However, there are terms that differ in the two amplitudes which are of order
$\lambda_C \simeq 1/4.5$, and so it is difficult to make 
a more precise quantitative estimate.

The decay rate of the neutron can be obtained from a general operator 
analysis \cite{wilczek}.  
For example, $\Gamma(p \rightarrow  \ell^+ \pi^0)/\Gamma(n \rightarrow
\ell^+ \pi^-) \simeq {1 \over 2}$ and $\Gamma(n \rightarrow 
\overline{\nu}K^0)/
\Gamma(p \rightarrow  \overline{\nu}K^+) \simeq 1.8$.  

{\bf 5. Doublet--triplet splitting and the standard $d=5$ proton decay:}
An important issue that any realistic GUT model faces is the question
of doublet--triplet splitting.  While the $SU(2)_L$ doublets in the
${\bf 10}_H$ of $SO(10)$ have to be light in order
to trigger electroweak symmetry breaking, their color triplet 
GUT partners have to remain heavy at the GUT
scale, since they mediate proton decay.  
One attractive feature of supersymmetric $SO(10)$ is the existence
of a natural doublet--triplet splitting mechanism \cite{dw, dwp}.  
This mechanism can bring in a numerical suppression in proton decay 
arising from the usual $({\bf 16}_i{\bf 16}_j){\bf 10}_H$ coupling 
that gives rise to the quark and lepton masses.  These operators, 
which lead to $p \rightarrow \overline{\nu} K^+$ as the dominant
mode, are somewhat problematic in supersymmetric $SU(5)$ since the
predicted rate is near the experimental limit \cite{murayama, nath}.  

The doublet--triplet splitting mechanism in $SO(10)$ utilizes the
superpotential couplings
\beq
W_{DT} = \Lambda {\bf 10}_H {\bf 45}_H {\bf 10}'_H + M' {\bf 10}_H'
{\bf 10}_H'~,
\eeq
where the ${\bf 10}_H$ is the Higgs superfield which contains the
two Higgs doublets of MSSM and where ${\bf 10}_H'$ is another
field with a GUT scale mass $M'$.  Once the ${\bf 45}_H$ acquires
a VEV along the $(B-L)$ direction, $\left \langle {\bf 45}_H \right
\rangle = {\rm diag}.(a,a,a,0,0) \otimes \tau_2$, the color triplet
mass matrix ${\cal M}$ and the $SU(2)$--doublet mass matrix ${\cal
M}'$ become 
\bea
{\cal M} = \left( \matrix{0 & \Lambda a \cr \Lambda a & M'} \right);
~~~~~~~~~~ {\cal M}' = \left(\matrix{0 & 0 \cr 0 & M'}\right)~.
\eea
This gives GUT scale masses to all the triplets from the
${\bf 10}_H$ and ${\bf 10}_H'$ while one pair of
Higgs doublets from ${\bf 10}_H$ remains light.  
Proton decay amplitude mediated by the color
triplets in ${\bf 10}_H$ is now proportional to $({\cal M}^{-1})_{11}
= (M'/\Lambda^2 a^2)$, so by choosing $M'$ somewhat smaller than
$\Lambda a$, one obtains a numerical suppression of proton decay.
$M'$ cannot be too small compared to the GUT scale, however, since that
would result in a large $positive$ contribution to the predicted value
of $\alpha_3(M_Z)$.  The shift in $\alpha_3(M_Z)$ from the doublet--triplet
sector alone is given by 
\beq \Delta \alpha_3(M_Z) = - {\alpha_3(M_Z)^2 \over 2 \pi}
{9 \over 7} {\rm ln}\left[({\cal M}^{-1})_{11} M_X \right]~.
\eeq
If $({\cal M}^{-1})_{11} = [10^{17}~GeV]^{-1}$, $\Delta \alpha_3 =
+0.005$, which might be acceptable.  If $({\cal M}^{-1})_{11} =
[10^{18}~GeV]^{-1}$ so that proton decay from this operator is
unobservable, then $\Delta \alpha_3(M_Z) \simeq 0.011$ which
seems excessive and would require a cancellation from some
other threshold effects.  This implies that
$p \rightarrow \overline{\nu} K^+$ cannot be suppressed
to an unobservable level, at least in this simple
doublet--triplet splitting scheme.  

In order for the VEV of ${\bf 45}_H$ to be 
along the $(B-L)$ direction to a great accuracy, the ${\bf 45}_H$ should not
couple or should couple only weakly to the ${\bf 16}_H+\overline{\bf 16}_H$
sector.  There is a danger of having pseudo--Goldstone multiplets,
which could upset the unification of the gauge couplings,
if such cross--couplings are prevented.  The potential
pseudo--Goldstones are the $\{(3,2,{1 \over 6}) + (3^*, 1, -{2 \over 3})
+ H.c.\}$ components, under $SU(3)_C \times SU(2)_L \times U(1)_Y$, 
from the ${\bf 16}_H$ and the ${\bf 45}_H$.  
This issue has been
addressed in the case of using ${\bf 16}_H + \overline{\bf 16}_H$
to break $SO(10)$ to $SU(5)$ \cite{dwp}.  Here we point out that the same
mechanism will work if the ${\bf 16}_H+\overline{\bf 16}_H$
is replaced by a ${\bf 126}_H+\overline{\bf 126}_H$.  The couplings
used in Eq. (6) that were relevant for the proton decay operators
are precisely the ones that can give masses to all these
would--be--Goldstone multiplets.  So in this case the proton
decay amplitude is closely related to the doublet--triplet
splitting mechanism.

{\bf 6. Gauge boson--mediated $d=6$ versus the new $d=5$ proton decay
operators:}  So far we have focussed on the $d=5$ proton decay
operators.  In the simplest supersymmetric GUT models, 
the gauge boson mediated $d=6$ operators
are suppressed relative to the $d=5$ operators.  However,
enhancement of the $d=6$ operators could occur
for a variety of reasons.  Possibility arises in
flipped $SU(5) \times U(1)$ \cite{ellis}, in
non--supersymmetric two--step breaking of $SO(10)$ models
\cite{parida}, or possibly even
in supersymmetric $SU(5)$ \cite{threshold} and $SO(10)$ models with 
large threshold corrections, wherein the
relevant $(X,Y)$ gauge boson masses are of order $10^{15}~GeV$.
Exchange of these particles 
would lead to a dominant $e^+ \pi^0$ mode, with proton
lifetime $\sim (10^{32}-10^{35})$ yrs., compatible with
current limits.  We wish to point out that should the $e^+ \pi^0$
decay mode of the proton be observed, one can empirically decide
whether it has its origin in the gauge--boson--mediated $d=6$ or in the
new $d=5$ operators discussed here.  For the gauge--mediated case,
$e^+ K^0$ will be strongly suppressed compared to the $e^+ \pi^0$
mode, by the Cabibbo angle (sin$^2\theta_C \simeq 1/20$), phase space
$(\simeq 1/2)$ and relevant matrix element-squared ($\simeq 1/2$), so
that $[\Gamma(e^+ K^0)/\Gamma(e^+ \pi^0)]_{d=6} \approx 1/80$.  By
contrast, for the new $d=5$ operators, we have shown that $e^+ K^0$
exceeds $e^+ \pi^0$ rate by about a factor of two.

{\bf 7.}  In conclusion the following remarks are in order.

(i) While the new $d=5$ proton decay operators seem to be best motivated
by their link to neutrino masses, we wish to note that 
the results presented here are more general.  Indeed, the
prominence of charged lepton modes and the predictions for certain
branching ratios depend only on the assumed non--hierarchical nature
of $\hat{F}$ and $\hat{G}$ in Eq. (10).  This condition might be
satisfied in other contexts as well.  For example, if the CKM angles
are induced in $SO(10)$ by coupling the fermions to two ${\bf 10}_H$
of Higgses, a reasonable fit may be obtained when the $SU(2)_L$
doublets and the color triplets from one of the ${\bf 10}_H$ has
non--hierarchical couplings, with strength of order the strange quark
Yukawa coupling.  The exchange of these color triplets would lead to
results similar to the ones presented here.  

(ii) As has been discussed in the literature, $d=5$ proton decay
operators such as $QQQL/M$, belonging to ${\bf 16}_i{\bf 16}_j
{\bf 16}_k {\bf 16}_l/M$, could be induced not only by the exchange
of GUT--related color triplets, but also by other effects
including exchange of
the heavy tower of color triplet string states.  These are
allowed by $SO(10)$ symmetry as well as as $R$--parity (or the $Z_2$
symmetry mentioned in Sec. 3).  In any supersymmetric theory, these
non--renormalizable operators must somehow be suppressed at least by a
factor of $10^{-7}$ (if $M \sim M_{\rm Pl}$) in order not to conflict
with observed limits on proton lifetime.  For a discussion of
this issue and its possible resolutions through the use of
flavor symmetries, in the context of
string--derived solutions, see Ref. \cite{patip}, and in a
non--string context, see for example, Ref. \cite{kaplan}.  

(iii) It is worth noting that there may be circumstances where 
the potentially dangerous GUT--related color triplets are projected
out, e.g., below the compactification scale of a string theory that
leads to a non-GUT symmetry like $G_{224} \subset SO(10)$ \cite{rizos},
but the components of ${\bf 16}_H$ and $\overline{\bf 16}_H$ providing
Majorana masses of the right--handed neutrinos may still exist below
the string scale.  In this case, the standard $d=5$ operators will be
absent, but the new $d=5$ operators discussed here could still be
effective.  

In summary, we have shown that in a class of grand unified theories
including supersymmetric $SO(10)$, there can be a significant link 
between the neutrino masses and proton decay.  In the
process of generating neutrino masses one typically induces 
a new source of $d=5$ proton decay interactions with an interesting
strength.  The flavor structure
of these new $d=5$ operators is distinctive.  In contrast to the
standard $d=5$ operators, the new ones
can lead to prominent (or even dominant) 
charged lepton decay modes, such as $\ell^+ \pi^0, \ell^+ K^0$ and
$\ell^+ \eta$, where $\ell = e$ or $\mu$, even for low or moderate
values of tan$\beta \stackrel{_<}{_\sim} 10$, along with
$\overline{\nu} K^+$ and $\overline{\nu} \pi^+$ modes.  A
distinguishing feature of the new mechanism, relative to
$d=6$ vector exchange, is the predicted ratio
$\Gamma(\ell^+ K^0):\Gamma(\ell^+ \pi^0) \simeq 2:1$. 

{\bf Acknowledgments:} K.S.B and F.W are supported by 
DOE grant No. DE-FG02-90ER-40542.  
J.C.P is supported in part by NSF Grant No. Phy-9119745 and by
a Distinguished Research Fellowship awarded by the University of
Maryland.  He acknowledges the hospitality and the stimulating
environment which he enjoyed during s short visit to the Institute
for Advanced Study.


\begin{thebibliography}{99}

\bibitem{pati}
        J. Pati and A. Salam, Phys. Rev. Lett. {\bf 31}, 661 (1973) and 
	Phys. Rev. {\bf D10}, 275 (1974);\\
        H. Georgi and S.L. Glashow, Phys. Rev. Lett. {\bf 32},
        438 (1974); \\
        H. Georgi, H. Quinn and S. Weinberg, Phys. Rev. Lett. {\bf
        33}, 451 (1974).  
     
\bibitem{nir}
        S. Dimopoulos, S. Raby and F. Wilczek, Phys. Rev. {\bf D24}, 
        1681 (1981);\\
	For recent reviews see for eg., P. Langacker and N. Polonsky, 
        Phys. Rev. {\bf D47}, 4028 (1993) and references therein.  

\bibitem{sakai}
        N. Sakai and T. Yanagida, Nucl. Phys. {\bf B197}, 533 (1982);\\
        S. Weinberg, Phys. Rev. {\bf D26}, 287 (1982).

\bibitem{raby}
        S. Dimopoulos, S. Raby and F. Wilczek, Phys. Lett. {\bf B112},
        133 (1982);\\
        J. Ellis, D.V. Nanopoulos, and S. Rudaz, Nucl. Phys. {\bf B202},
        43 (1982);
        
\bibitem{murayama}
        J. Hisano, H. Murayama and T. Yanagida, Nucl Phys. {\bf B402},
        46 (1993).

\bibitem{babubarr}
	K.S. Babu and S.M. Barr, Phys. Lett. {\bf B381}, 137 (1996).

\bibitem{nath}
        P. Nath and R. Arnowitt, hep-ph/9708469.

\bibitem{so10}
        H. Georgi,, in {\it Particles and Fields} 1974, ed.
        C.E. Carlson, (AIP, NY, 1975) p. 575;\\
        H. Fritzsch and P. Minkowski, Ann. Phys. {\bf 93}, 193 (1975).

\bibitem{seesaw}
        M. Gell-Mann, P. Ramond and R. Slansky, in: $Supergravity$, eds.
        F. van Nieuwenhuizen and D. Freedman (Amsterdam, North Holland, 1979) 
        p. 315; \\
        T. Yanagida, in: {\it Workshop on the Unified Theory and Baryon Number
        in the Universe}, eds. O. Sawada and A. Sugamoto (KEK, Tsukuba) 95 
        (1979); \\
        R.N. Mohapatra and G. Senjanovic, Phys. Rev. Lett. {\bf 44}, 912 
        (1980).

\bibitem{bl}
        See for eg., R.N. Mohapatra, Phys. Rev. {\bf D34}, 3457 (1986);\\
        A. Font, L. Ibanez and F. Quevedo, Phys. Lett. {\bf B288}, 
        79 (1989);\\
        S.P. Martin, Phys. Rev. {\bf D46}, 2769 (1992).

\bibitem{gj}
        H. Georgi and C. Jarlskog, Phys. Lett. {\bf 86B}, 297 (1979).

\bibitem{bbfermion}
        K.S. Babu and S.M. Barr, Phys. Rev. {\bf D56}, 2614 (1997). 

\bibitem{jarlskog}
        C. Jarlskog, Phys. Lett. {\bf 82B}, 401 (1979).

\bibitem{wilczek}
        S. Weinberg, Phys. Rev. Lett. {\bf 43}, 1566 (1979);\\
        F. Wilczek and A. Zee, Phys. Rev. Lett. {\bf 43}, 1569 (1979).

\bibitem{dw}
        S. Dimopoulos and F. Wilczek, Report No. NSF-ITP-82-07, 1981,
	in {\it The unity of fundamental interactions}, Proceedings of
        the 19th Course of the International School on Subnuclear
        Physics, Erice, Italy, 1981, ed. by A. Zichichi (Plenum Press,
        New York, 1983). 

\bibitem{dwp}
        K.S. Babu and S.M. Barr, Phys. Rev. {\bf D48}, 5354 (1993);
        {\bf D50}, 3529 (1994); {\bf D51}, 2463 (1995);\\
        S.M. Barr and S. Raby, hep-ph/9705336.

\bibitem{ellis}
        J. Ellis, J. Lopez and D.V. Nanopoulos,  Phys. Lett.
	{\bf B371}, 65 (1996);\\
        S.M. Barr, Phys. Lett. {\bf 112B}, 219 (1982).

\bibitem{parida}
        For recent updates see for eg., D. Lee, R.N. Mohapatra, 
        M.K. Parida and M. Rani, Phys. Rev. {\bf D51}, 229 (1995) and
        references therein.  

\bibitem{threshold}
	J. Hisano, Y. Nomura and T. Yanagida, hep-ph/9710279.

\bibitem{patip}
        J.C. Pati, Phys. Lett. {\bf B388}, 532 (1996) and hep-ph/9611371, 
        Proc. Intl. Workshop on
        {\it Future prospects of baryon instability}, ed. by
        Y. Kamyshkov, Oak Ridge, March 28-30, (1996); \\
        A. Farraggi, Nucl. Phys. {\bf B428}, 111 (1994).

\bibitem{kaplan}
	H. Murayama and D. Kaplan, Phys. Lett. {\bf B336}, 221 (1994); \\
        C. Carone, L. Hall and H. Murayama, Phys. Rev. {\bf D53}, 
        6282 (1996).
 
\bibitem{rizos}
        I. Antoniadis, G. Leontaris and J. Rizos, Phys. Lett. {\bf
        B245}, 161 (1990).
  
\end{thebibliography}
\end{document}